\documentclass[conference]{IEEEtran}
\usepackage[letterpaper, top=0.75in, bottom=0.95in, left=0.63in, right=0.63in]{geometry}
\usepackage[nolist]{acronym}
\usepackage{float}
\usepackage{fontawesome}
\usepackage{amsmath}
\usepackage{tikz}

\usepackage{graphicx}
\usepackage{subcaption}
\usepackage{xcolor}     
\usepackage{tabularray}  
\let\labelindent\relax
\usepackage{enumitem}
\usepackage[T1]{fontenc}
\usepackage{url}
\usepackage{booktabs}
\usepackage{multirow}
\usetikzlibrary{arrows.meta,positioning,calc,shadows}

\definecolor{HeaderSoftBlue}{HTML}{D9EAF7}
\definecolor{OurSoftGray}{HTML}{F2F2F2}

\ifCLASSINFOpdf
\else
\fi
\begin{document}
\acrodef{3GPP}{3rd Generation Partnership Project}
\acrodef{5G}{5th Generation Mobile Network}
\acrodef{6G}{6th Generation Mobile Network}
\acrodef{AI}{Artificial Intelligence}
\acrodef{AI4Net}{\ac{AI} for Networking}
\acrodef{MDP}{Markov Decision Process}
\acrodef{AIDER}{Aerial Image Dataset for Emergency Response}
\acrodef{AMF}{Access and Mobility Management Function}
\acrodef{AIaaS}{Artificial Intelligence-as-a-Service}
\acrodef{AC}{Actor-Critic}
\acrodef{AIDER}{Aerial Image Database for Emergency Response applications}
\acrodef{B5G}{Beyond Fifth Generation}
\acrodef{BPF}{Berkeley Packet Filter}
\acrodef{CBR}{Constant Bit Rate}
\acrodef{CSV}{Comma-Separated Values}
\acrodef{CPU}{Central Processing Unit}
\acrodef{CNN}{Convolutional Neural Network}
\acrodef{CNNs}{Convolutional Neural Networks}
\acrodef{C-V2X}{Cellular Vehicle to-Everything}
\acrodef{DoS}{Denial of Service}
\acrodef{DDQL}{Double Deep
 Q-learning}
\acrodef{DDoS}{Distributed Denial of Service}
\acrodef{DDPG}{Deep Deterministic Policy Gradient}
\acrodef{DNN}{Deep Neural Network}
\acrodef{DRL}{Deep Reinforcement Learning}
\acrodef{DQN}{Deep Q-Network}
\acrodef{DT}{Decision Tree}
\acrodef{DDQN}{Double Deep Q-Network}

\acrodef{ETSI}{European Telecommunications Standards Institute}
\acrodef{eNWDAF}{Evolved Network Data Analytics Function}
\acrodef{eBPF}{Extended Berkeley Packet Filter}
\acrodef{ECDF}{Empirical Cumulative Distribution Function}
\acrodef{ECDFs}{Empirical Cumulative Distribution Functions}
\acrodef{FIBRE}{Future Internet Brazilian Environment for Experimentation}
\acrodef{FL}{Federated Learning}
\acrodef{FEL}{Federated Ensemble Learning}
\acrodef{GNN}{Graph Neural Networks}
\acrodef{GPU}{Graphics Processing Unit}
\acrodef{GTP}{GPRS Tunnelling Protocol}
\acrodef{GTP-U}{GPRS Tunnelling Protocol User Plane}
\acrodef{GA}{Genetic Algorithm}
\acrodef{HTM}{Hierarchical Temporal Memory}

\acrodef{IAM}{Identity And Access Management}
\acrodef{ICMP}{Internet Control Message Protocol}
\acrodef{IID}{Independent and Identically Distributed}
\acrodef{IoE}{Internet of Everything}
\acrodef{IoT}{Internet of Things}
\acrodef{ITU}{International Telecommunication Union}
\acrodef{IQR}{Interquartile Range}
\acrodef{I/O}{Input/Output}
\acrodef{IP}{Internet Protocol}
\acrodef{KNN}{K-Nearest Neighbors}
\acrodef{KPI}{Key Performance Indicator}
\acrodef{KPIs}{Key Performance Indicators}
\acrodef{LSTM}{Long Short-Term Memory}
\acrodef{LOWESS}{Locally Weighted Scatterplot Smoothing}
\acrodef{LR}{Learning Rate}
\acrodef{LLM}{Large Language Model}
\acrodef{MAE}{Mean Absolute Error}
\acrodef{MAD}{Median Absolute Deviation}
\acrodef{ML}{Machine Learning}
\acrodef{MLaaS}{Machine Learning as a Service}
\acrodef{MOS}{Mean Opinion Score}
\acrodef{MAPE}{Mean Absolute Percentage Error}
\acrodef{MSE}{Mean Squared Error}
\acrodef{MEC}{Multi-access Edge Computing}
\acrodef{mMTC}{Massive Machine Type Communications}
\acrodef{MFA}{Multi-factor Authentication}
\acrodef{MLP}{Multi-Layer Perceptron}
\acrodef{MADRL}{Multi-Agent Deep Reinforcement Learning}
\acrodef{MAB}{Multi-Armed Bandit}
\acrodef{MILP}{Mixed Integer Linear Programming}
\acrodef{MQTT}{Message Queuing Telemetry Transport}
\acrodef{NWDAF}{Network Data Analytics Function}
\acrodef{Net4AI}{Networking for \ac{AI}}
\acrodef{NS}{Network Slicing}
\acrodef{NFV}{Network Function Virtualization}
\acrodef{NEB}{Net Energy Benefit}
\acrodef{OSM}{Open Source MANO}
\acrodef{PCA}{Principal Component Analysis}
\acrodef{PC-FedAvg}{Personalized Conditional Federated Averaging}
\acrodef{PoC}{Proof of Concept}
\acrodef{PPO}{Proximal Policy Optimization}
\acrodef{POMDP}{Partially Observable Markov decision process}
\acrodef{PCAP}{Packet Capture}
\acrodef{PSO}{Particle Swarm Optimization}
\acrodef{PDU}{Power Distribution Unit}
\acrodef{QoE}{Quality of experience}
\acrodef{QoS}{Quality of Service}
\acrodef{QFI}{QoS Flow Identifier}
\acrodef{QFIs}{QoS Flow Identifiers}
\acrodef{RAM}{Random Access Memory}
\acrodef{RF}{Random Forest}
\acrodef{RL}{Reinforcement Learning}
\acrodef{RMSE}{Root Mean Square Error}
\acrodef{RNN}{Recurrent Neural Network}
\acrodef{RTT}{Round-Trip Time}
\acrodef{RAN}{Radio Access Network}
\acrodef{RTP}{Real-time Transport Protocol}
\acrodef{RAPL}{Running Average Power Limit}
\acrodef{SDN}{Software-Defined Networking}
\acrodef{SFI2}{Slicing Future Internet Infrastructures}
\acrodef{SLA}{Service-Level Agreement}
\acrodef{SON}{Self-Organizing Network}
\acrodef{SMF}{Session Management Function}
\acrodef{S-NSSAI}{Single Network Slice Selection Assistance Information}
\acrodef{SVM}{Support Vector Machine}
\acrodef{SOPS}{Service-Aware Optimal
 Path Selection}
 \acrodef{SAFE}{Scalable Asynchronous Federated Ensembling}
\acrodef{TQFL}{Trust Deep Q-learning Federated Learning}
\acrodef{TEID}{Tunnel Endpoint Identifier}
\acrodef{TEIDs}{Tunnel Endpoint Identifiers}
\acrodef{TPE}{Tree-Structured Parzen Estimator}
\acrodef{UE}{User Equipment}
\acrodef{UEs}{User Equipments}
\acrodef{UPF}{User Plane Function}
\acrodef{UPFs}{User Plane Functions}
\acrodef{URLLC}{Ultra-Reliable and Low Latency Communications}
\acrodef{UAV}{Unmanned Aerial Vehicle}
\acrodef{UAVs}{Unmanned Aerial Vehicles}
\acrodef{UDP}{User Datagram Protocol}
\acrodef{VoD}{Video on Demand}
\acrodef{VR}{Virtual Reality}
\acrodef{AR}{Augmented Reality}
\acrodef{V2V}{Vehicle-to-Vechile}
\acrodef{V2X}{Vehicle-to-Everything}
\acrodef{VNF}{Virtual Network Function}
\acrodef{VNFs}{Virtual Network Functions}
\acrodef{VPN}{Virtual Private Network}
\acrodef{XDP}{eXpress Data Path}
%


\title{Energy-Aware Agentic Orchestration\\for Multi-Site Service Deployment}

\author{\IEEEauthorblockN{
Rodrigo Moreira\textsuperscript{1}, Alisson Chaves\textsuperscript{2, }\textsuperscript{3}, Larissa F. {Rodrigues Moreira}\textsuperscript{1}, Tereza Carvalho\textsuperscript{4},\\ Tiago Barros\textsuperscript{3}, Daniel Corujo\textsuperscript{3}, Rui Aguiar\textsuperscript{3}, David Gomez-Barquero\textsuperscript{5}, Flávio de Oliveira Silva\textsuperscript{6}}
\IEEEauthorblockA{
\textsuperscript{1}Federal University of Viçosa (UFV), Minas Gerais, Brazil\\
\textsuperscript{2}Federal University of Uberlândia (UFU), Minas Gerais, Brazil\\
\textsuperscript{3}Instituto de Telecomunicações and Universidade de Aveiro (UA), Aveiro, Portugal\\
\textsuperscript{4}University of São Paulo (USP), São Paulo, Brazil\\
\textsuperscript{5}Universitat Politecnica de València (UPV)
Valencia, Spain\\
\textsuperscript{6}University of Minho (UMinho), Braga, Portugal\\
Emails: \{rodrigo, larissa.f.rodrigues\}@ufv.br, alisson@ufu.br, terezacarvalho@usp.br,\\ \{aoc.alisson, tiagobarros, dcorujo, ruilaa\}@av.it.pt, dagobar@iteam.upv.es, flavio@di.uminho.pt}
}


%


\maketitle

\begingroup
\renewcommand\thefootnote{}
\footnotetext{\scriptsize
© 2026 IEEE. Personal use of this material is permitted. Permission from IEEE must be obtained for all other uses, in any current or future media, including reprinting/republishing this material for advertising or promotional purposes, creating new collective works, for resale or redistribution to servers or lists, or reuse of any copyrighted component of this work in other works.

Accepted author manuscript for publication in the Proceedings of the 2026 IEEE Future Networks World Forum (FNWF 2026), Lisbon, Portugal, October 26--28, 2026.
}
\addtocounter{footnote}{-1}
\endgroup

\begin{abstract}
Energy-aware orchestration across geographically distributed infrastructures is increasingly important as data centers absorb growing Artificial Intelligence (AI) workloads. However, existing approaches often exclude the energy cost of automated decision-making from system-level sustainability assessments. This paper presents an energy-aware agentic orchestration framework that treats Large Language Model (LLM) reasoning as a measurable component of the control loop. We compare single-shot tool calling, ReAct, and plan-and-execute strategies on a multi-site Kubernetes testbed spanning two European data centers and instrumented at layers ranging from power distribution units to per-pod energy attribution. Using ten natural-language orchestration intents with varying decision complexity, we show that reasoning strategy and intent complexity strongly influence graphics processing unit (GPU) energy consumption during orchestration decisions. ReAct and plan-and-execute generated executable tool calls across all evaluated intents but required multiple LLM inferences, whereas the single-shot configuration was less costly but non-actionable. Prefix caching reduced mean ReAct GPU energy consumption by 29.4\%. For the evaluated lightweight workload, the infrastructure-energy changes observed within the one-second measurement window were insufficient to offset the measured GPU decision energy. These findings demonstrate the importance of accounting for reasoning overhead when assessing the sustainability of AI-native orchestration and motivate energy-budgeted selection of agent strategies.
\end{abstract}


\IEEEpeerreviewmaketitle

\section{Introduction}\label{sec:introduction}

Digital services have become the connective tissue of contemporary economies, sustaining domains as diverse as public administration, industrial automation, and consumer entertainment. As these services become more interactive and data-intensive, providers increasingly distribute their workloads across multiple geographic sites, making multi-site container orchestration the operational substrate of modern cloud and edge platforms. According to the International Energy Agency (IEA), data centers already account for close to 1.5\% of global electricity demand and are projected to more than double their consumption by 2030, a trajectory amplified by the rapid uptake of accelerator-bound artificial intelligence workloads. Deciding when, where, and how to deploy containerized services across geographically distributed infrastructures without compromising sustainability has thus emerged as a central, unresolved operational challenge~\cite{Yang2025, Danach2026, Andersen2026, Moreira2026}.

Two decades of work in cluster scheduling, autoscaling, and energy-aware placement have produced rich heuristic, optimization, and learning-based controllers, and modern Kubernetes ecosystems already expose layered telemetry from power distribution units down to per-pod attribution~\cite{Kocot2023, Jawaddi2024, Mohamed2025}. More recently, the operations community has begun to investigate \ac{LLM} agents as a new orchestration interface, translating natural-language intents into sequences of monitoring and actuation calls through patterns such as single-shot tool calling, ReAct, and plan-and-execute. The promise is substantial: operators could request the most energy-efficient site for a new workload in plain language, get causal explanations for sudden power spikes, or trigger graceful migrations during congestion events, thereby indirectly benefiting end users with greener, more reliable services. 

However, existing studies evaluate these agents primarily by task accuracy and operational effectiveness, while generally treating the energy consumed by the decision-making process as negligible relative to the managed infrastructure. This assumption may not hold when iterative reasoning executes repeatedly on accelerator-based platforms. In energy-aware orchestration, excluding this control-plane overhead can overestimate the sustainability benefits attributed to an automated decision. Determining whether, and under which conditions, the operational value of \ac{LLM}-driven orchestration justifies its additional energy cost therefore remains an important open question~\cite{Lee2026, Bellal2026}.

To address this gap, this paper proposes an energy-aware agentic framework for orchestrating containerized services across geographically distributed Kubernetes clusters, in which the energy consumed by the agent itself is treated as a first-class, measurable quantity alongside the energy consumed by the infrastructure it influences. The framework couples a reasoning agent with a typed tool interface that exposes fine-grained \ac{PDU}, \ac{RAPL}, and per-pod telemetry as well as lifecycle actuation across sites, and instantiates the same pipeline under three reasoning strategies evaluated on a benchmark of ten natural-language intents stratified by decision complexity.  We hypothesize that the sustainability of agentic orchestration depends on the interaction between intent complexity, reasoning strategy, and workload magnitude: more adaptive strategies improve actionability at the cost of additional inference energy, while lightweight workloads may fail to amortize this reasoning cost.

Our key contributions are: (i) an energy-aware agentic orchestration framework incorporating LLM decision energy into multi-site infrastructure assessment; (ii) the \ac{NEB} metric comparing short-horizon infrastructure savings against accelerator decision energy; (iii) a real two-site Kubernetes testbed instrumented from PDUs down to per-pod attribution and GPU accelerators; and (iv) an empirical evaluation quantifying how intent complexity, reasoning strategy, and prefix caching impact decision cost, latency, and actionability.

Sections~\ref{sec:related_work}, \ref{sec:method}, \ref{sec:testbed}, and~\ref{sec:results} present the related work, method, testbed, and results, respectively; Section~\ref{sec:concluding_remarks} concludes the paper.


\section{Related Work}\label{sec:related_work}

This section reviews energy-aware orchestration approaches across edge, cloud, wireless, and \ac{AI}-serving systems, highlighting how prior work has addressed energy efficiency, and situates our agentic approach within this landscape.

\begin{table*}[htbp]
\centering
\caption{Qualitative comparison of energy-aware orchestration approaches.}
\label{tab:related-work}
\scriptsize
\begin{tblr}{
  width = 0.9\textwidth,
  colspec = {
    X[l,m,2.4]  
    X[c,m,0.8]  
    X[c,m,0.8]  
    X[c,m,0.8]  
    X[c,m,0.8]  
    X[c,m,0.8]  
    X[c,m,0.8]  
    X[c,m,0.8]  
    X[c,m,0.8]  
    X[c,m,0.8]  
  },
  row{1} = {font=\bfseries, bg=HeaderSoftBlue},
  row{12} = {font=\bfseries, bg=OurSoftGray},
  hline{1,2,12,Z} = {-}{},
  hline{12} = {0.6pt},
  rows = {rowsep=2pt},
  columns = {colsep=4pt},
}
Approach & Energy & QoS & Telemetry & Runtime & Multi-site & Tools/API & NL agent & Agent Cost & Real Eval. \\
Munir, S. et al.~\cite{paper1}  & \faCircle & \faCircleO  & \faCircleO  & \faCircle & \faCircleO  & \faCircleO  & \faCircleO  & \faCircleO  & \faCircle \\
Premsankar, G. et al.~\cite{paper2}  & \faCircle & \faCircle & \faCircleO  & \faCircle & \faCircleO  & \faCircleO  & \faCircleO  & \faCircleO  & \faCircle \\
Aguirre-Usandizaga, J. et al.~\cite{paper3}  & \faCircle & \faCircleO  & \faCircle & \faCircleO  & \faCircleO  & \faCircleO  & \faCircleO  & \faCircleO  & \faCircle \\
Kohli, H. et al.~\cite{paper4}  & \faCircle & \faCircle & \faCircle & \faCircle & \faCircleO  & \faCircleO  & \faCircleO  & \faCircleO  & \faCircleO  \\
Kadhim, M. et al.~\cite{paper5}  & \faCircle & \faCircle & \faCircle & \faCircle & \faCircleO  & \faCircleO  & \faCircleO  & \faCircleO  & \faCircleO  \\
Suryadevara, S.~\cite{paper6}  & \faCircle & \faCircle & \faCircle & \faCircle & \faCircle & \faCircle & \faCircleO  & \faCircleO  & \faCircle \\
Zawish, M. et al.~\cite{paper7}  & \faCircle & \faCircleO  & \faCircle & \faCircleO  & \faCircleO  & \faCircleO  & \faCircleO  & \faCircleO  & \faCircleO  \\
Vance, E.~\cite{paper8}  & \faCircle & \faCircle & \faCircleO  & \faCircleO  & \faCircleO  & \faCircleO  & \faCircleO  & \faCircleO  & \faCircleO  \\
Maisto, V. et al.~\cite{paper9}  & \faCircle & \faCircle & \faCircle & \faCircle & \faCircleO  & \faCircleO  & \faCircleO  & \faCircleO  & \faCircle \\
Moore, H. et al.~\cite{paper10} & \faCircle & \faCircle & \faCircleO  & \faCircle & \faCircle & \faCircleO  & \faCircleO  & \faCircleO  & \faCircle \\
\textbf{Our} &
\faCircle &
\faCircle &
\faCircle &
\faCircle &
\faCircle &
\faCircle &
\faCircle &
\faCircle &
\faCircle \\
\end{tblr}
\end{table*}

Prior work addresses energy-aware orchestration through reinforcement learning, optimization, forecasting, and heuristic scheduling across MEC dispatch, DNN placement, GPU power prediction, sustainable AI workload management, wireless resource coordination, and carbon-aware cloud orchestration~\cite{paper1,paper2,paper3,paper4,paper5,paper6}. Related efforts consider edge-AI pruning, green-cloud optimization, multi-tenant DNN inference, and sustainable LLM serving across geographically distributed data centers~\cite{paper7,paper8,paper9,paper10}.


Table~\ref{tab:related-work} compares these approaches across key operational dimensions. Existing methods cover subsets of these metrics but generally treat the controller's decision-making cost as external to the system budget. Our method combines telemetry-informed, tool-enabled multi-site orchestration with explicit measurement of agentic decision overhead.

\section{Proposed Method}\label{sec:method}

\begin{figure}[htbp]
    \centering
    \includegraphics[width=0.99\columnwidth]{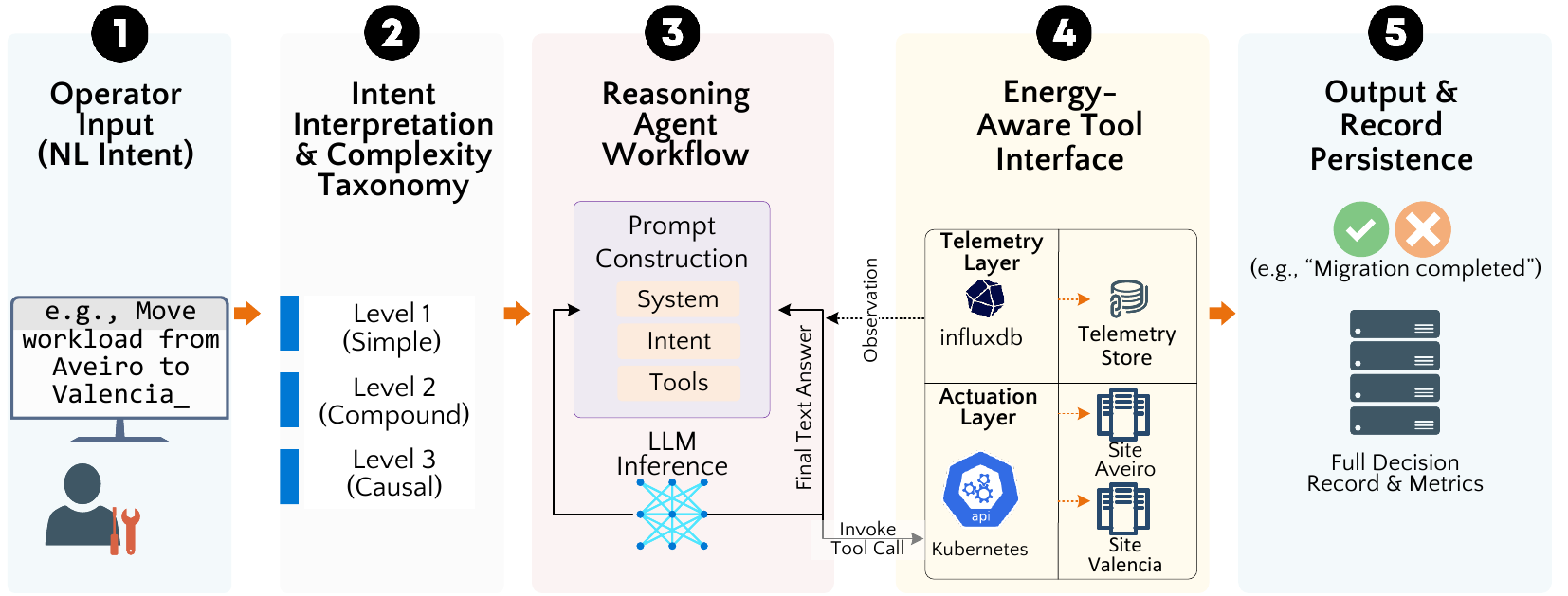}
    \caption{End-to-end pipeline of the proposed method.}
    \label{fig:method-overview}
\end{figure}

Figure~\ref{fig:method-overview} summarizes the proposed energy-aware agentic framework, which maps natural-language intents to orchestration decisions using current site-energy telemetry.


The method comprises four coupled components. The first is an intent interpretation layer that receives the operator's natural-language request and routes it to a reasoning agent. The second is a telemetry access layer, which exposes fine-grained energy measurements of the infrastructure as callable tools. The third is an actuation layer that exposes lifecycle operations on target deployments as callable tools. The fourth is the reasoning agent itself, whose strategy determines how it orchestrates these tools to reach a decision. We describe each component and then explain how they combine into the end-to-end workflow shown in Figure~\ref{fig:method-overview}.

Each intent is a free-form request that refers to one or more sites and expresses a goal related to energy, performance, or placement. Requests range from simple observations, such as querying a site's current power consumption, to complex optimization goals involving multiple sites and constraints. We categorize intents by complexity level based on the number and nature of operations required to satisfy them. Level~1 intents are satisfiable by a single query or a single actuation. Level~2 intents require a sequence of reads and writes with an explicit dependency between them. Level~3 intents require causal reasoning, hypothetical projection, or multi-constraint optimization across sites. This taxonomy is not a property of the natural language itself but of the decision process that the intent demands. It lets us analyze how different reasoning strategies behave as decision complexity grows.

The agent interacts with the infrastructure exclusively through a typed tool interface. Two families of tools are exposed. The first family grants read access to the infrastructure's energy state. It includes tools to retrieve aggregate site power, per-pod energy-consumption ranked by magnitude, host-level power measured at the power distribution unit, energy time series for trend analysis, and a cross-site comparison that enables evidence-based placement decisions. The second family grants the orchestrator write access. It includes tools to query deployment status, scale deployments, migrate workloads between sites, and operate a deployment in both sites simultaneously.

Each tool is described to the agent using a natural-language signature that specifies its purpose and parameters. The agent chooses which tools to call, in what order, and with what arguments. This separation is important for the analysis that follows. The telemetry tools are side-effect-free and can be called repeatedly at low infrastructural cost. Actuation tools modify the cluster's live state and can only be evaluated by executing them. The energy cost of the agent's reasoning, as opposed to the cost of the tools themselves, is therefore a distinct and measurable quantity.


The first is a single-shot tool-use baseline, in which the model receives the intent together with the tool catalog and is expected to emit the required tool calls in a single generation.

The second is an iterative strategy based on the ReAct paradigm, in which the agent alternates between reasoning steps and tool invocations. After each tool call, the agent observes the result and decides whether to continue reasoning, call another tool, or terminate with a final answer. This strategy generates multiple inference requests per intent, and the number of requests increases with decision complexity. It offers explicit self-correction at the cost of additional inferences.

The third is a plan-and-execute strategy, in which the agent first emits a complete sequence of tool calls based solely on the intent and the tool catalog, and then executes the sequence without further reasoning. This strategy separates planning from execution. In our implementation, it may require additional model calls for plan normalization or final response synthesis, but it does not perform the observation-reasoning loop used by ReAct.

These strategies occupy distinct positions on the space defined by reasoning overhead and adaptivity. The experimental evaluation in Section~\ref{sec:results} measures where each strategy lands in practice.

\begin{figure}[htbp]
    \centering
    \includegraphics[width=0.99\columnwidth]{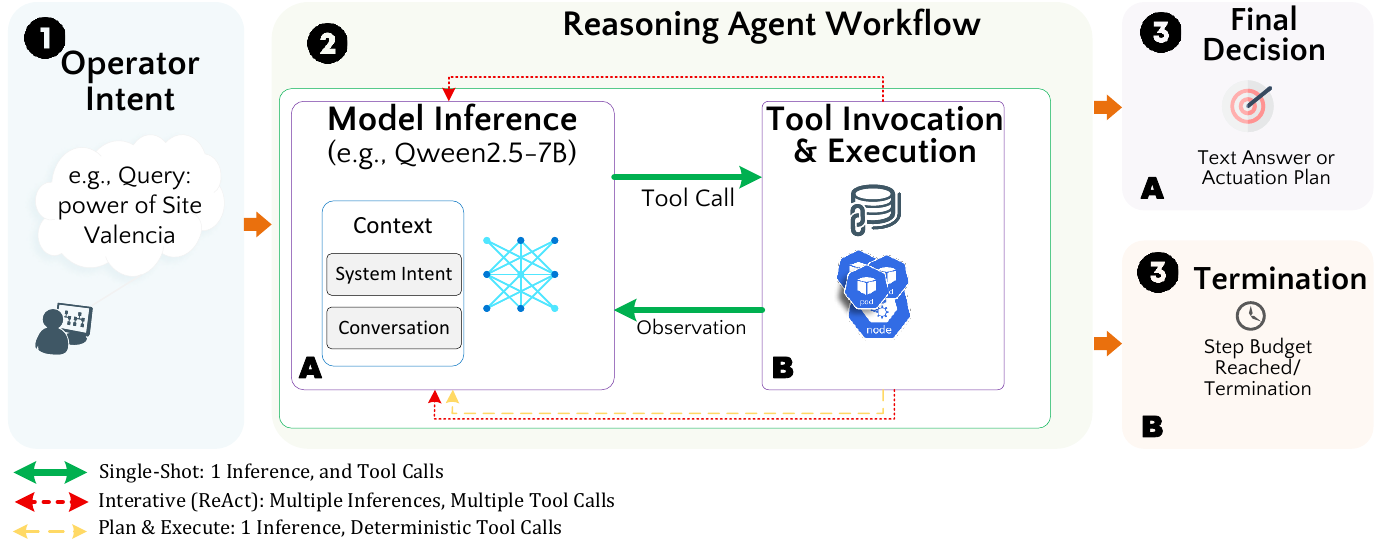}
    \caption{Agent reasoning loop.}
    \label{fig:reasoning-flow}
\end{figure}

Figure~\ref{fig:reasoning-flow} details the internal loop that the agent executes when processing an intent. The agent first combines the intent with a system prompt that describes the available tools and the orchestration context. The resulting prompt is submitted to the language model, which produces either a textual response, one or more tool calls, or both. When the model emits tool calls, it dispatches each call to the corresponding telemetry or actuation tool. The tool result is appended to the conversation and, if the strategy permits, fed back to the model for the next reasoning step. The loop terminates when the model emits a final textual answer, when a strategy-specific step budget is exhausted, or when a tool returns a terminal condition.

A key observation is that the single-shot strategy performs only the first inference and terminates after the initial response, regardless of whether it emits valid tool calls.

Because the method treats reasoning cost as a measurable quantity, we introduce an explicit attribution scheme that separates three contributions to the total energy associated with an intent. We quantify energy consumption across three layers. We measure inference energy at the compute accelerator hosting the language model. Managed workload energy is measured at the power distribution unit and attributed at each site's software layer. We estimate the energy impact of each agent action from the change in infrastructure energy consumption observed before and after the intervention over a fixed measurement window.

For action intents, the observed short-horizon infrastructure-energy difference is computed as $E_{\mathrm{saved}}=(P_{\mathrm{before}}-P_{\mathrm{after}})\Delta t$, where $P_{\mathrm{before}}$ and $P_{\mathrm{after}}$ are the mean cluster powers measured over one-second windows immediately before and after the action, respectively, and $\Delta t=1$ second. We then define the net energy benefit as $NEB=E_{\mathrm{saved}}-E_{\mathrm{GPU}}$, where $E_{\mathrm{GPU}}$ is the accelerator energy consumed during the complete decision window. A positive \ac{NEB} indicates that the observed infrastructure-energy reduction exceeded the decision energy within the evaluated observation window. A negative value indicates that the decision energy was not recovered within that window; it does not establish whether the action could be energetically amortized over a longer operational period.


The end-to-end workflow, illustrated in Figure~\ref{fig:method-overview}, proceeds as follows. The system receives an intent and pairs it with the system prompt. The agent enters the reasoning loop described above, calling telemetry tools to gather evidence about the sites' energy state and calling actuation tools when a decision requires changing the live cluster. Throughout this process, the compute accelerator energy is sampled at fixed intervals and integrated over time to obtain the intent's reasoning cost. Before and after any actuation, the monitoring layer records the infrastructure energy state to enable the attribution described above. Once the agent produces its final answer, the \ac{NEB} is computed for every action-oriented benchmark intent from the paired infrastructure measurements and the corresponding decision energy. Configurations that produced no actuation are retained as non-action baselines, and the full decision record is persisted for analysis.


This workflow is strategy agnostic. The same pipeline serves all three reasoning strategies by varying only the step budget and the prompt instructions provided to the agent. This property is what makes the comparative evaluation in Section~\ref{sec:results} possible: the reasoning strategy is the independent variable, while the rest of the system is held constant.

\section{Experimental Testbed}\label{sec:testbed}

\begin{figure}[t]
    \centering
    \includegraphics[width=0.90\columnwidth]{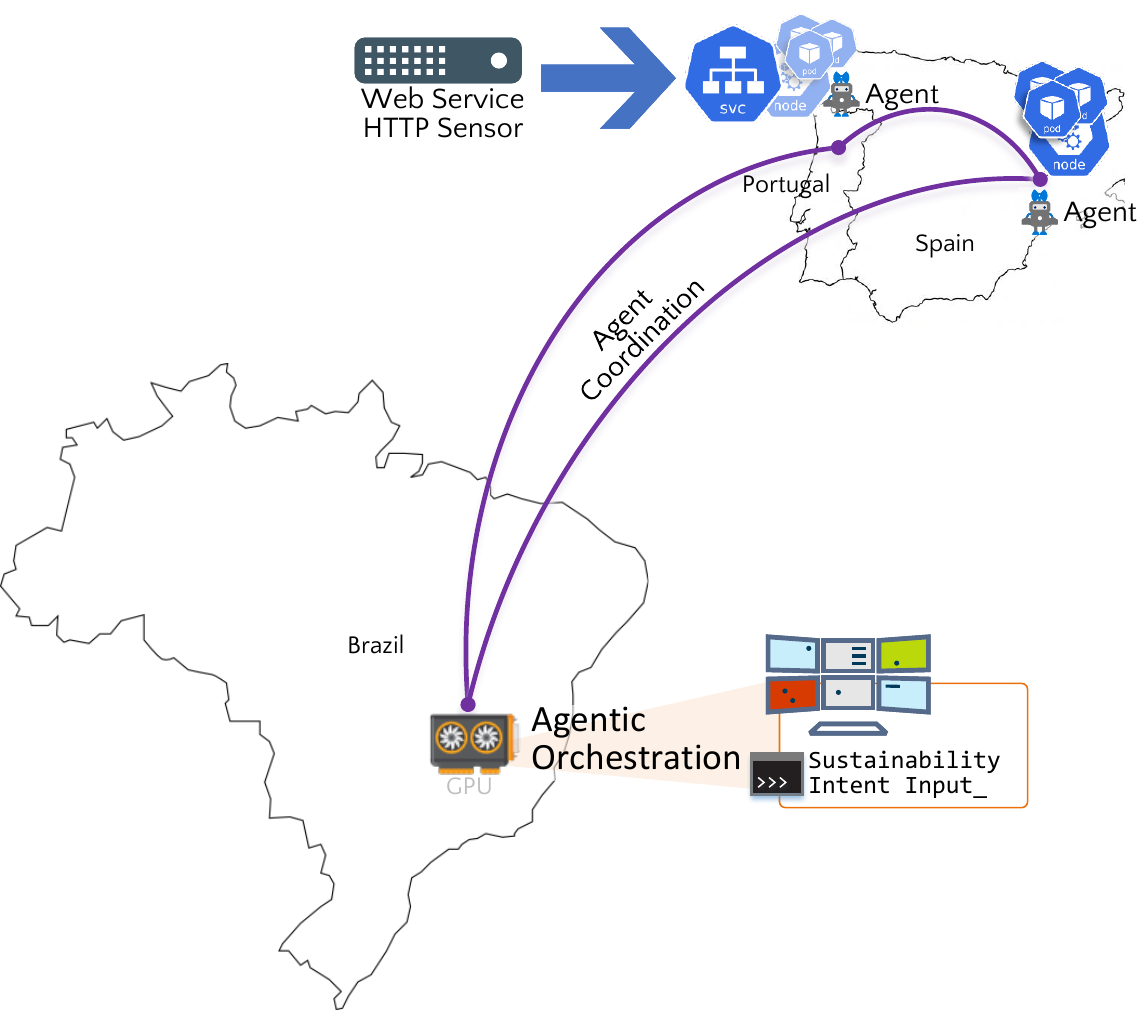}
    \caption{Three-tier experimental topology.}
    \label{fig:testbed}
\end{figure}

We evaluate the proposed method on a two-site Kubernetes deployment spanning two European data centers, with the reasoning agent hosted at a third geographic location. Figure~\ref{fig:testbed} depicts the overall topology and the flow of telemetry and orchestration traffic across the three sites. Table~\ref{tab:intents} summarizes the benchmark intents used in the evaluation.

\begin{table}[t]
\centering
\caption{Benchmark of ten natural language intents used to evaluate the proposed method.}
\label{tab:intents}

\begin{tblr}{
  width = \linewidth,
  colspec = {
    X[c,m,0.7]
    X[c,m,0.9]
    X[l,m,3]
    X[c,m,0.9]
  },
  row{1} = {font=\bfseries, bg=HeaderSoftBlue},
  column{1} = {font=\bfseries}, 
  column{3} = {halign=l},
  hline{1,2,5,9,Z} = {-}{},
  hline{5,9} = {0.6pt},
  rows = {rowsep=3pt},
  columns = {colsep=5pt},
}
Level & ID & Natural Language Intent & Action \\

\SetCell[r=3]{c,m,font=\scriptsize} L1
  & L1\_01 & What is the current power consumption of site Aveiro?                                      & No  \\
  & L1\_02 & Scale the web service deployment to 3 replicas in Valencia.                                & Yes \\
  & L1\_03 & Which pod is consuming the most energy in Aveiro right now?                                & No  \\

\SetCell[r=4]{c,m,font=\scriptsize} L2
  & L2\_04 & Reduce energy consumption in Aveiro by 20\%.                                              & Yes \\
  & L2\_05 & Move the workload from Aveiro to Valencia.                                                & Yes \\
  & L2\_06 & Balance energy consumption evenly across both sites.                                      & Yes \\
  & L2\_07 & Which site is more energy-efficient per pod right now?                                    & No  \\

\SetCell[r=3]{c,m,font=\scriptsize} L3
  & L3\_08 & Energy in Aveiro spiked in the last 5 minutes. What happened?                             & No  \\
  & L3\_09 & If I deploy 5 more replicas of web service, which site would have lower total energy cost?  & No  \\
  & L3\_10 & Optimize overall energy across both sites without reducing total replicas.                & Yes \\
\end{tblr}

\end{table}

The agent and language model run on a workstation in Brazil equipped with an Intel Xeon w3-2435 CPU (8 cores, 30\,GiB RAM) and an NVIDIA RTX 4000 Ada GPU (20\,GiB VRAM, 130\,W TDP). The GPU serves Qwen2.5-7B-Instruct via vLLM~0.12.0, with the agent built on LangChain~1.2.0 and LangGraph~1.0.5.


The infrastructure consists of a single Kubernetes~1.28.15 cluster split across two sites connected by a \ac{VPN} tunnel: the primary site at ITAV (Aveiro, Portugal) hosting 3 control plane and 3 worker nodes, and a secondary compute site at UPV (Valencia, Spain) with 3 worker nodes. Control plane responsibilities reside at Aveiro, while Valencia acts as a compute pool.


ITAV nodes run Ubuntu~24.04 on kernel~6.8 with containerd as the container runtime. UPV nodes run Ubuntu~22.04 on kernel~5.15 with the same runtime. The cluster uses Flannel for pod networking over VXLAN. The sites are reachable via private addresses: 10.255.40.0/24 for Aveiro and 172.21.32.0/24 for Valencia, and traffic between them traverses the \ac{VPN} tunnel that connects the two networks.

The workload under orchestration is a lightweight distributed web application deployed as a Kubernetes Deployment with no explicit resource requests or limits. The agent scales, migrates, and duplicates replicas of this deployment across the two sites. Because the sites host no competing workloads during the experiments, the application's contribution to each worker's power envelope is small relative to the host's baseline consumption. This property is relevant to the interpretation of the \ac{NEB} reported in Section~\ref{sec:results}.

Each worker node is instrumented to measure energy at three levels, which together provide a layered view of consumption from the physical outlet down to the process. At the physical layer, a PDU measures the bare-metal power drawn by each host. At the operating system layer, Intel RAPL exposes per-socket energy counters for the CPU package and its memory controller. At the container orchestration layer, Kepler runs as a per-node DaemonSet in a dedicated namespace and attributes energy to individual pods and containers by combining RAPL counters with cgroup-level resource accounting. All three data sources are published to a single InfluxDB instance at one-second granularity, and the agent consumes them via the Holosense platform, which exposes the underlying store as a query API. The instrumentation covers both Aveiro and Valencia workers symmetrically, enabling the agent to compare the two sites on an equivalent basis.

The power draw of the \ac{GPU} hosting the language model is sampled at 1\,Hz through the NVIDIA Management Library, and the resulting power trace is integrated using the trapezoidal rule to estimate the \ac{GPU} energy consumed per intent. The integration spans the full decision window, including \ac{LLM} inference, tool execution, and communication delays, and therefore includes the accelerator's power draw during these activities. It does not include the energy consumed by the agent host's CPU, memory, or network interface; consequently, $E_{\mathrm{GPU}}$ represents accelerator-side decision energy rather than the total control-plane energy of orchestration.


The evaluation comprises two phases that differ only in their prefix-cache configuration. Phase~1 uses the FP16 model with vLLM's prefix cache enabled, whereas Phase~2 uses the same model and precision with the prefix cache disabled. In each phase, all 10 benchmark intents are evaluated under each of the three reasoning strategies, with five repetitions per intent--strategy combination executed in randomized order. Before every action intent, the cluster is automatically restored to a canonical baseline in which Aveiro hosts one replica and Valencia hosts none, thereby providing a comparable initial placement state across runs. A 5-second cooldown is inserted between consecutive runs to reduce short-term thermal carryover on the \ac{GPU}.


Every run produces a record that combines four sources of evidence. The \ac{GPU} energy trace is produced locally on the agent host. The inference-level telemetry, which includes prompt and generation token counts, as well as the prefix cache hit rate, is obtained from vLLM's Prometheus endpoint and differenced over the run window. The cluster energy state, sampled at both the PDU and Kepler layers, is retrieved from Holosense immediately before and after any action intent. The agent's orchestration loop records the reasoning trace, which captures the sequence of model inferences and tool invocations. The four sources are aligned on a common timeline and persisted in a single CSV file per phase for subsequent statistical analysis.

\section{Results and Discussion}\label{sec:results}

This section evaluates how the reasoning strategy, intent complexity, prefix caching, and workload magnitude, relative to the cluster baseline, shape the energy cost, operational effectiveness, and net benefit of agentic orchestration.

\begin{figure}[htbp]
    \centering
    \includegraphics[width=0.9\columnwidth]{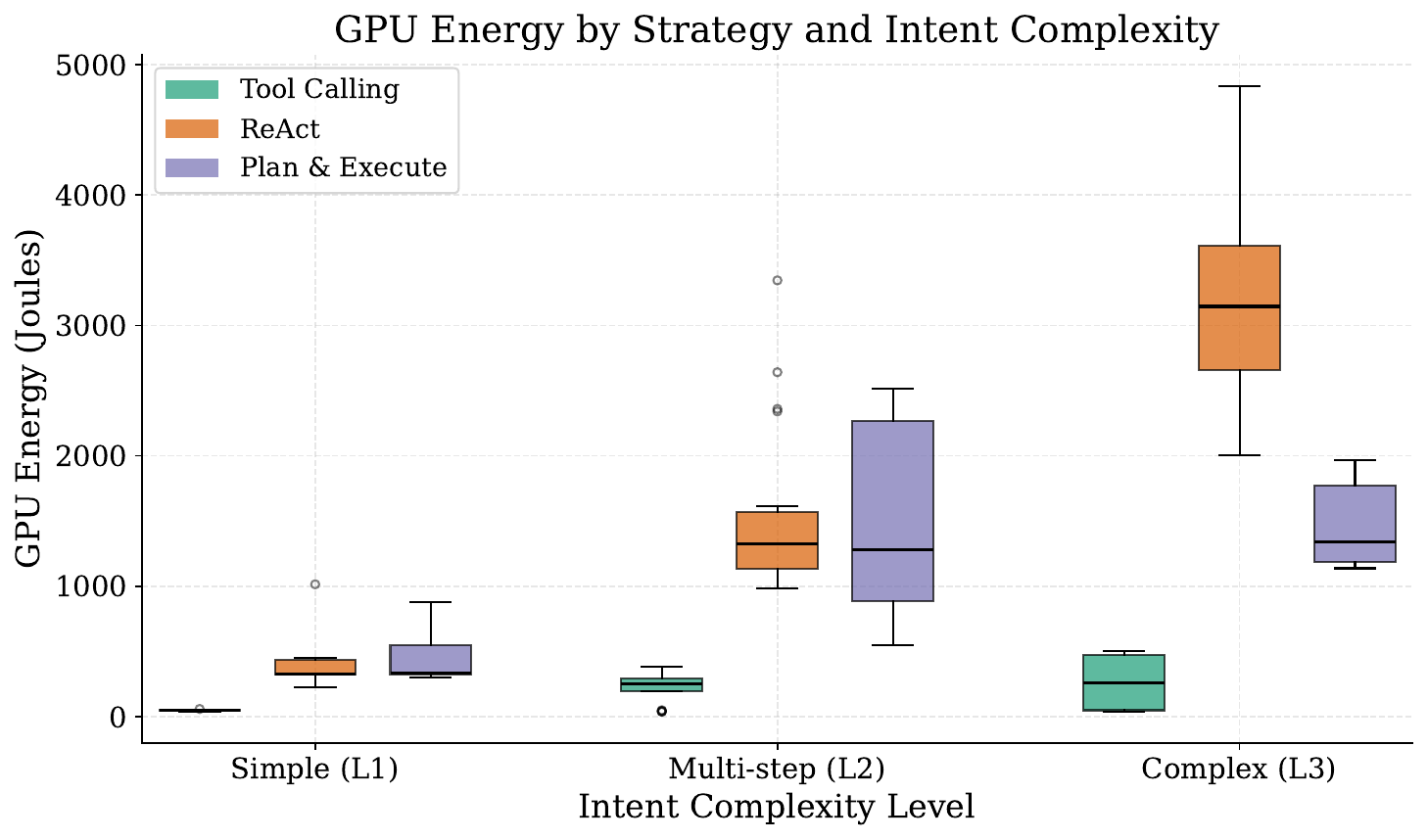}
    \caption{\ac{GPU} energy by reasoning configuration and intent complexity with prefix caching enabled. ReAct consumed approximately 8--12$\times$ the energy of the non-actionable single-shot Tool~Calling baseline.}
    \label{fig:energy-by-strategy}
\end{figure}

\textbf{Energy Cost Across Reasoning Configurations}. Figure~\ref{fig:energy-by-strategy} and Table~\ref{tab:summary} compare mean \ac{GPU} energy across the three reasoning configurations and complexity levels under the cache-enabled condition. For Tool~Calling, mean energy increased from 48\,J for simple intents to 264\,J for complex intents. ReAct increased from 395\,J to 3{,}277\,J, corresponding to 8.3$\times$ and 12.4$\times$ the energy of the non-actionable single-shot baseline, respectively. Plan~\&~Execute ranged from 449\,J to 1{,}476\,J and consumed substantially less energy than ReAct for complex intents. These measurements quantify computational cost; the actionability of the configurations is examined separately below.


The observed energy differences relate to the number of \ac{LLM} inference calls required per intent. Table~\ref{tab:summary} shows that Tool~Calling consistently performs one inference call regardless of complexity, whereas ReAct increases from 2.0 calls for simple intents to 5.3 for complex intents, and Plan~\&~Execute increases from 2.0 to 3.0 calls. Each additional inference extends the decision window and processes newly generated tokens within an expanding context. Although prefix caching can reuse shared portions of this context, it does not eliminate the energy cost of subsequent inference steps. The number of calls and accumulated context therefore help explain the higher energy consumption of the iterative configurations.


\begin{table}[htbp]
\centering
\caption{Mean \ac{GPU} energy ($\pm$ SD), \ac{LLM} and tool calls, and latency by reasoning configuration and intent complexity with prefix caching enabled. Five repetitions per intent--configuration pair; $n$ is pooled by complexity level.}
\label{tab:summary}
\small

\resizebox{\columnwidth}{!}{
\begin{tabular}{llrrrrr}
\toprule
\textbf{Strategy} & \textbf{Level} & \textbf{$n$} & \textbf{Energy (J)} & \textbf{LLM} & \textbf{Tools} & \textbf{Lat. (s)} \\
\midrule
Tool Calling & L1 & 15 & 48 $\pm$ 5 & 1.0 & 0.0 & 1.6 \\
             & L2 & 20 & 233 $\pm$ 124 & 1.0 & 0.0 & 3.5 \\
             & L3 & 15 & 264 $\pm$ 187 & 1.0 & 0.0 & 3.8 \\
\midrule
ReAct        & L1 & 15 & 395 $\pm$ 183 & 2.0 & 1.0 & 6.6 \\
             & L2 & 20 & 1{,}532 $\pm$ 643 & 3.1 & 2.6 & 20.9 \\
             & L3 & 15 & 3{,}277 $\pm$ 817 & 5.3 & 5.8 & 36.8 \\
\midrule
Plan \& Exec & L1 & 15 & 449 $\pm$ 199 & 2.0 & 1.0 & 6.8 \\
             & L2 & 20 & 1{,}450 $\pm$ 747 & 2.5 & 2.5 & 19.9 \\
             & L3 & 15 & 1{,}476 $\pm$ 308 & 3.0 & 4.3 & 16.3 \\
\bottomrule
\end{tabular}
}

\end{table}

\begin{figure}[htbp]
    \centering
    \includegraphics[width=0.9\columnwidth]{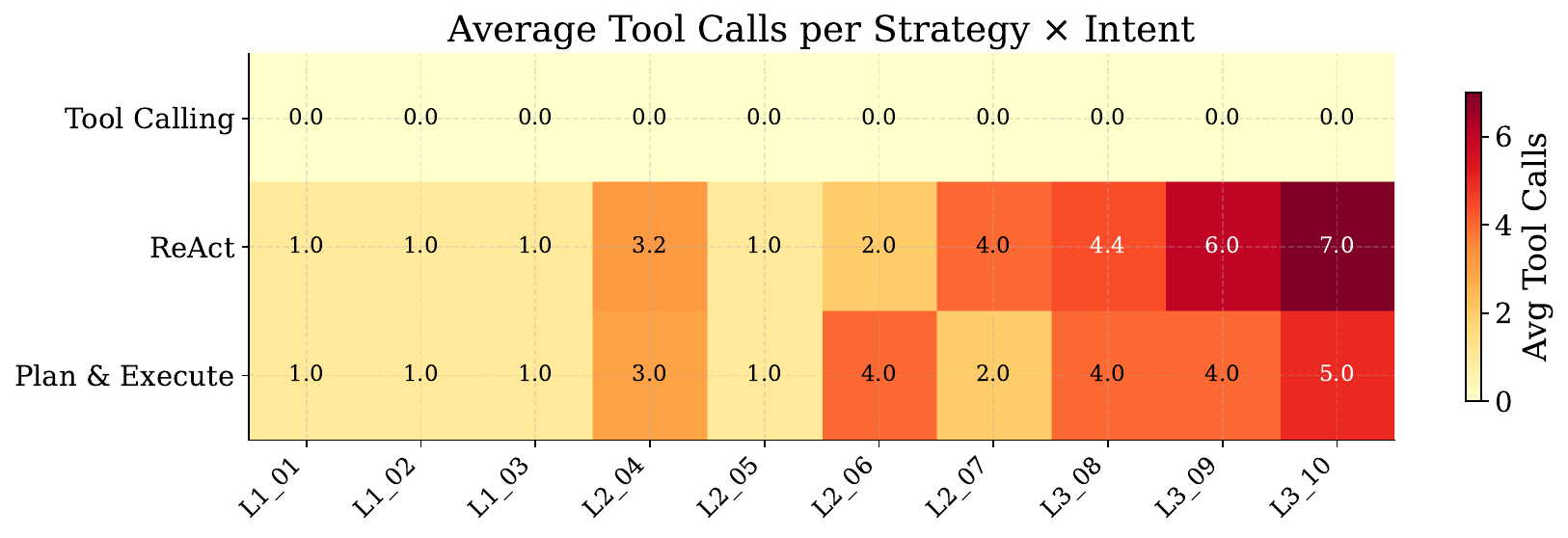}
    \caption{Average tool calls per strategy and intent.}
    \label{fig:tool-heatmap}
\end{figure}

\textbf{Single-Shot Tool Calling: Low Cost but Non-Actionable}. Table~\ref{tab:summary} and Figure~\ref{fig:tool-heatmap} show that the single-shot configuration produced no tool invocations across the 100 evaluated runs. Under the adopted prompt and Qwen2.5-7B-Instruct configuration, the model generated textual responses instead of structured tool calls. Consequently, this configuration neither retrieved infrastructure telemetry nor executed Kubernetes actions. Its energy measurements therefore represent a low-cost, non-actionable generation baseline rather than an operationally equivalent orchestration strategy. This result is specific to the evaluated model and configuration because tool-use behavior may depend on the model, prompt, chat template, and structured-output implementation.


\begin{figure}[htbp]
    \centering
    \includegraphics[width=0.9\columnwidth]{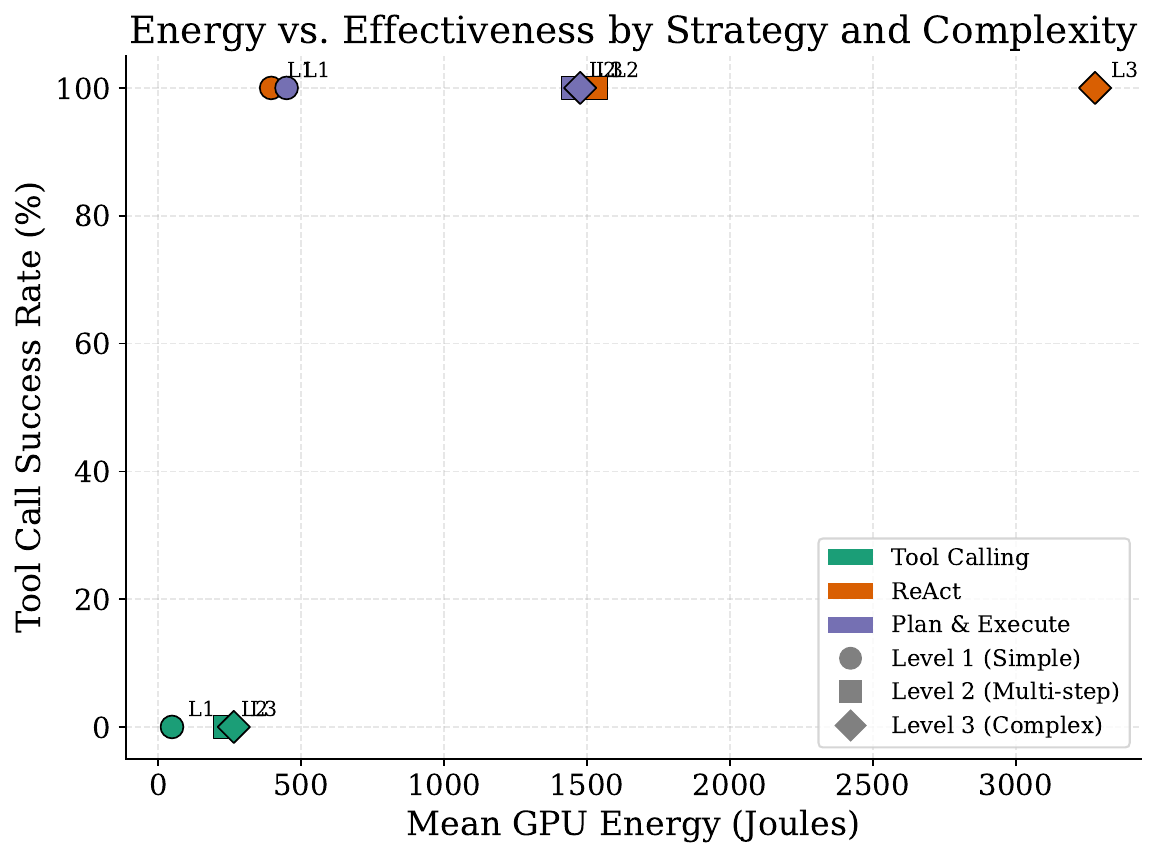}
    \caption{GPU energy versus tool-call actionability rate.}
    \label{fig:correctness}
\end{figure}

Figure~\ref{fig:correctness} illustrates the energy--actionability trade-off observed in the experiment. ReAct and Plan~\&~Execute produced executable tool calls across all evaluated complexity levels, corresponding to a 100\% tool-call actionability rate, whereas the single-shot configuration produced none. Because the single-shot configuration completed no operational task, its lower energy consumption should not be interpreted as greater orchestration efficiency. In this setup, the two actionable strategies consumed approximately 8--12 times as much GPU energy as the non-actionable baseline, demonstrating that the energy cost of achieving actionability must be accounted for explicitly. The reported rate indicates only whether a strategy produced executable tool calls; it does not assess the semantic correctness or global optimality of the resulting orchestration decision.


\begin{table}[htbp]
\centering
\caption{Statistical analysis of \ac{GPU} energy (Joules) across Cache and Complexity factors.}
\label{tab:robust-stats}
\tiny
\begin{tblr}{
  width = \columnwidth,
  colspec = {
    X[l,m,1.2]   
    X[l,m,1.5]   
    X[l,m,1.0]   
    X[c,m,1.1]   
    X[c,m,1.2]   
    X[c,m,1.0]   
  },
  row{1} = {font=\bfseries, bg=HeaderSoftBlue},
  row{2,10} = {font=\itshape, bg=black!5},
  hline{1,2,3,10,11,Z} = {-}{},
  hline{6,8,14,17} = {2-6}{0.5pt},
  rows = {rowsep=2pt},
  columns = {colsep=2pt},
}
Strategy & Test & Factor & Statistic & \textbf{$p$-value (Sig.)} & Effect Size \\

\SetCell[c=6]{l}
\textit{Assumptions: Shapiro-Wilk OK ($p=0.103$), Levene VIOLATED ($p<0.001$)} \\

\SetCell[r=7]{l,m} ReAct
  & ANOVA (Type II) & Cache       & $F=15.17$  & $2.6\times10^{-4}$   & $\eta^2=0.053$ \\
  &                 & Complexity  & $F=211.25$ & $1.2\times10^{-20}$  & $\eta^2=0.744$ \\
  &                 & Interaction & $F=1.48$   & 0.229 (ns)           & $\eta^2=0.005$ \\
  & Welch $t$       & Cache       & $t=-1.81$  & 0.076 (ns)           & $d=-0.48$ \\
  &                 & Complexity  & $t=-12.99$ & $<10^{-6}$           & $d=-3.41$ \\
  & Welch $t$ (L1)  & Cache       & $t=-3.53$  & 0.003                & $d=1.34$ \\
  & Welch $t$ (L3)  & Cache       & $t=-2.76$  & 0.011                & $d=1.04$ \\

\SetCell[c=6]{l}
\textit{Assumptions: Shapiro-Wilk VIOLATED ($p<0.001$), Levene VIOLATED ($p<0.001$)} \\

\SetCell[r=8]{l,m} \shortstack[l]{Tool\\Calling}
  & ANOVA (Type II)  & Cache       & $F=0.43$   & 0.516 (ns)           & $\eta^2=0.004$ \\
  &                  & Complexity  & $F=59.53$  & $2.3\times10^{-10}$  & $\eta^2=0.511$ \\
  &                  & Interaction & $F=0.43$   & 0.513 (ns)           & $\eta^2=0.004$ \\
  & Kruskal-Wallis   & All groups  & $H=29.36$  & $<10^{-5}$           & --- \\
  & Mann-Whitney $U$ & Cache       & $U=411$    & 0.569 (ns)           & $r=0.09$ \\
  &                  & Complexity  & $U=92$     & $<10^{-6}$           & $r=0.80$ \\
  & Welch $t$ (L1)   & Cache       & $t=0.11$   & 0.916 (ns)           & $d=-0.04$ \\
  & Welch $t$ (L3)   & Cache       & $t=-0.66$  & 0.517 (ns)           & $d=0.25$ \\
\end{tblr}
\end{table}

\begin{table}[t]
\centering
\caption{Pairwise complexity comparisons (Simple vs.\ Complex) per cache condition.}
\label{tab:pairwise-complexity}
\small
\resizebox{\columnwidth}{!}{
\begin{tabular}{llrrrl}
\toprule
\textbf{Strategy} & \textbf{Condition} & \textbf{Simple (J)} & \textbf{Complex (J)} & \textbf{Ratio} & \textbf{$p$ (Welch)} \\
\midrule
\multirow{2}{*}{ReAct}
  & Cache ON  & 395  & 3{,}277 & 8.3$\times$ & $<10^{-6}$ \\
  & Cache OFF & 975  & 4{,}383 & 4.5$\times$ & $<10^{-6}$  \\
\midrule
\multirow{2}{*}{Tool Calling}
  & Cache ON  & 48   & 264     & 5.5$\times$ & $<0.001$  \\
  & Cache OFF & 48   & 304     & 6.4$\times$ & $<10^{-5}$  \\
\bottomrule
\end{tabular}
}
\end{table}

\textbf{Statistical Validation}. The assumptions of the standard two-way ANOVA were not uniformly satisfied: homogeneity of variance was violated for both ReAct and Tool~Calling, and residual normality was additionally violated for Tool~Calling. We therefore complemented the ANOVA with Welch's $t$-test and the non-parametric Kruskal-Wallis and Mann-Whitney $U$ tests, as reported in Table~\ref{tab:robust-stats}. For the ReAct and Tool~Calling configurations included in this analysis, intent complexity exhibited the largest reported effect on \ac{GPU} energy. Table~\ref{tab:pairwise-complexity} further shows that complex intents consumed 4.5--8.3 times more energy than simple intents across the evaluated cache conditions.



\begin{figure}[htbp]
    \centering
    \includegraphics[width=0.9\columnwidth]{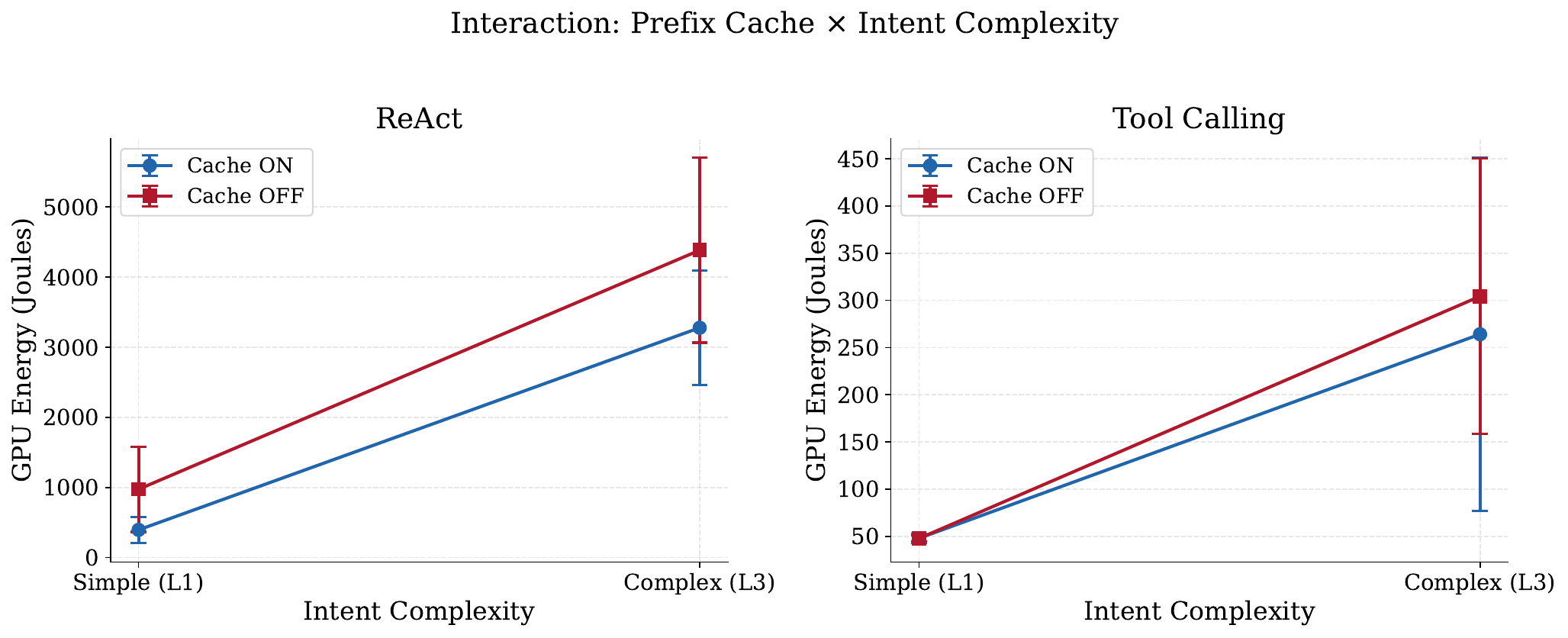}
    \caption{Effect of prefix caching and intent complexity on \ac{GPU} energy.}
    \label{fig:interaction-plot}
\end{figure}

\textbf{Strategy-Dependent Effect of Prefix Caching}. Figure~\ref{fig:interaction-plot} shows that enabling prefix caching reduced mean ReAct \ac{GPU} energy by 29.4\%, with reductions of 59.5\%, 32.8\%, and 25.2\% for levels L1, L2, and L3, respectively. The pooled ANOVA identified a significant cache effect, whereas Welch's test was marginal under heterogeneous variance; level-specific comparisons remained significant for L1 and L3. For Tool~Calling, the observed mean reduction was 9.3\% and was not statistically significant. This contrast is consistent with ReAct's repeated inference calls reusing a growing shared prefix, whereas the single-shot configuration provides no accumulated prefix reuse across inference steps.



\begin{figure}[htbp]
    \centering
    \includegraphics[width=0.9\columnwidth]{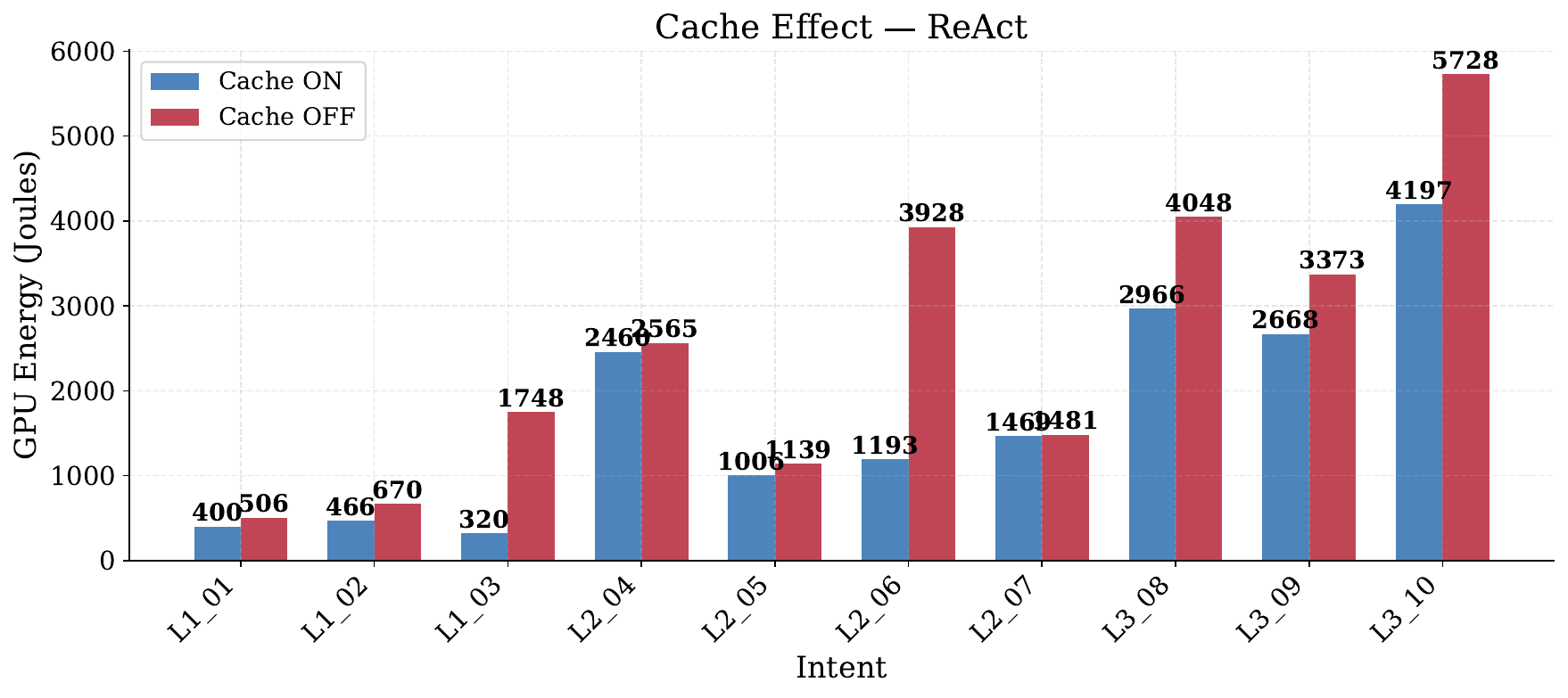}
    \caption{Per-intent ReAct \ac{GPU} energy with prefix caching enabled (blue) and disabled (red).}
    \label{fig:cache-react}
\end{figure}

Figure~\ref{fig:cache-react} compares per-intent ReAct energy with prefix caching enabled and disabled. The largest absolute reduction occurred for the complex intent L3\_10, where energy decreased from 5{,}728\,J to 4{,}197\,J. This pattern is consistent with key-value (KV) prefix caching: successive \ac{LLM} calls in a ReAct loop reuse a prefix containing the system prompt, tool definitions, and prior reasoning steps, allowing cached key-value representations to be reused rather than recomputed. Longer reasoning traces provide more opportunities for such reuse, although the resulting energy reduction also depends on context composition and runtime implementation.


\begin{figure}[t]
    \centering
    \includegraphics[width=0.99\columnwidth]{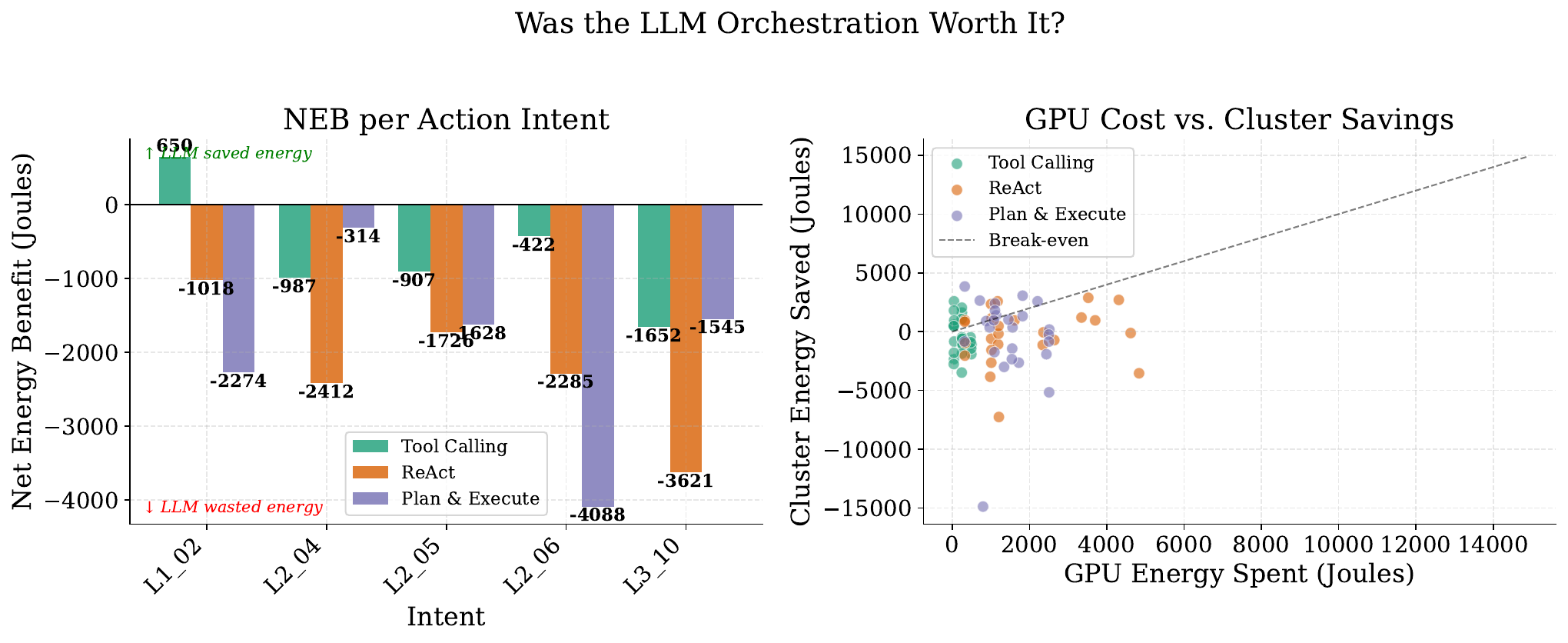}
    \caption{Short-horizon \ac{NEB} by action-oriented intent (left) and \ac{GPU} decision energy versus observed infrastructure-energy difference (right).}
    \label{fig:neb}
\end{figure}

\textbf{Short-Horizon Infrastructure Impact and \ac{NEB}}. Figure~\ref{fig:neb} reports the \ac{NEB} for the five action-oriented benchmark intents. Consistent with the definition in Section~\ref{sec:method}, $NEB=E_{\mathrm{saved}}-E_{\mathrm{GPU}}$, where $E_{\mathrm{saved}}$ is the infrastructure-energy difference observed across the one-second pre- and post-action windows and $E_{\mathrm{GPU}}$ is the accelerator energy consumed during the decision. The resulting values therefore characterize the short-horizon energy balance of each evaluated configuration, rather than the total savings that an action might accumulate over its operational lifetime.


Table~\ref{tab:neb} reports a negative \ac{NEB} for 14 of the 15 strategy--intent combinations under the evaluated one-second attribution horizon. The sole positive value, obtained for Tool~Calling on L1\_02, cannot be attributed to orchestration because this configuration produced no tool calls and executed no infrastructure action. Instead, it reflects short-term variation between the two one-second cluster-power measurement windows used to compute $E_{\mathrm{saved}}$. The Tool~Calling entries should therefore be interpreted as a non-action baseline that illustrates the sensitivity of short-window \ac{NEB} estimates to background power fluctuations.


\begin{table}[htbt]
\centering
\caption{Short-horizon \ac{NEB} for action-oriented intents with prefix caching enabled.}
\label{tab:neb}
\small

\begin{tblr}{
  width = \linewidth,
  colspec = {X[l,m,1.6] X[c,m,1.0] X[c,m,0.9] X[c,m,0.9] X[c,m,0.9]},
  row{1} = {font=\bfseries},        
  cells = {font=\scriptsize},       
  rows = {rowsep=0.5pt},              
  columns = {colsep=4pt},
  hline{1} = {0.2pt},                 
  hline{2} = {0.6pt},               
  hline{Z} = {1pt},                 
}
\textbf{Strategy} & \textbf{Intent} & \textbf{GPU (J)} & \textbf{Saved (J)} & \textbf{NEB (J)} \\

Tool Calling & L1\_02 & 46    & 696      & \textbf{+650} \\
             & L2\_04 & 255   & $-$732   & $-$987 \\
             & L2\_05 & 44    & $-$864   & $-$908 \\
             & L2\_06 & 254   & $-$168   & $-$422 \\
             & L3\_10 & 488   & $-$1{,}164 & $-$1{,}652 \\

\hline[0.6pt]

ReAct        & L1\_02 & 466   & $-$552   & $-$1{,}018 \\
             & L2\_04 & 2{,}460 & 48    & $-$2{,}412 \\
             & L2\_05 & 1{,}006 & $-$720 & $-$1{,}726 \\
             & L2\_06 & 1{,}193 & $-$1{,}092 & $-$2{,}285 \\
             & L3\_10 & 4{,}197 & 576   & $-$3{,}621 \\

\hline[0.6pt]

Plan \& Exec & L1\_02 & 606   & $-$1{,}668 & $-$2{,}274 \\
             & L2\_04 & 1{,}070 & 756   & $-$314 \\
             & L2\_05 & 1{,}664 & 36    & $-$1{,}628 \\
             & L2\_06 & 2{,}492 & $-$1{,}596 & $-$4{,}088 \\
             & L3\_10 & 1{,}569 & 24    & $-$1{,}545 \\
\end{tblr}

\end{table}

\begin{figure}[htbp]
    \centering
    \includegraphics[width=\columnwidth]{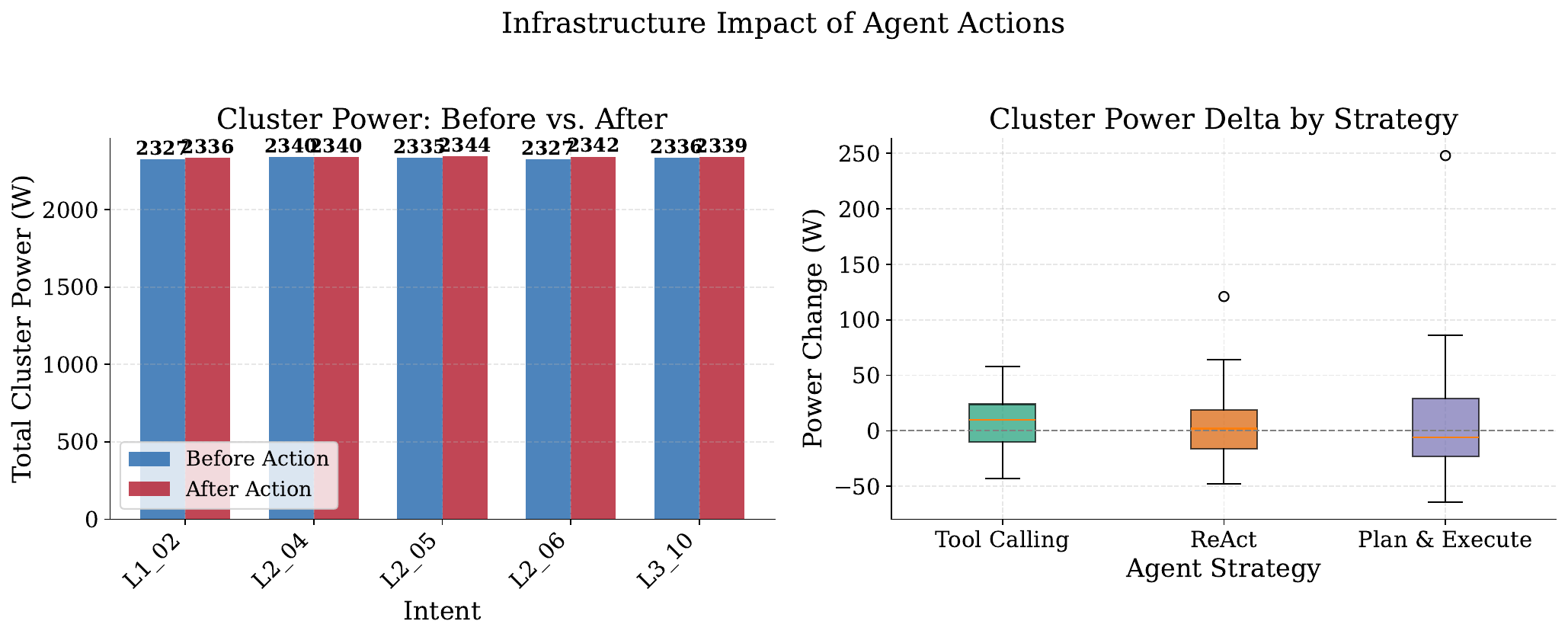}
     \caption{Cluster \ac{PDU} power before and after each orchestration attempt (left) and corresponding power difference by reasoning configuration (right).}
    \label{fig:cluster-power}
\end{figure}

Figure~\ref{fig:cluster-power} shows no consistent reduction in total cluster power between the one-second pre- and post-action observation windows. Because the lightweight web-service workload contributes only a small fraction of each host's overall power draw, any action-induced change is difficult to distinguish from short-term variation in the infrastructure baseline. Consequently, the observed infrastructure-energy differences were generally insufficient to offset the GPU energy consumed by LLM reasoning within the evaluated observation window.


Kubernetes execution records confirmed that the actionable strategies completed the intended operations. ReAct scaled down replicas for L2\_04 and migrated the workload for L2\_05, while Plan~\&~Execute completed the migration for L2\_05. The predominantly negative \ac{NEB} values should therefore not be interpreted as evidence that these actions failed to execute. Rather, under the evaluated lightweight workload and one-second observation windows, their power effects were not consistently distinguishable from short-term variation in the cluster baseline at the \ac{PDU} level. This result is specific to the evaluated workload and observation horizon and does not rule out a positive \ac{NEB} for larger workloads or longer operational horizons.


\textbf{Implications}. Three main observations follow from these results. First, \ac{LLM} reasoning introduces a measurable \ac{GPU} energy cost ($\sim$3{,}000\,J for complex ReAct intents) that must be explicitly budgeted in sustainable orchestration. Second, prefix caching provides a key optimization for iterative reasoning, reducing mean ReAct \ac{GPU} energy by 29.4\%. Third, negative \ac{NEB} values reflect our lightweight workload and 1-second observation window; larger workloads or longer observation horizons may amortize decision overhead and achieve net positive energy gains.

\section{Concluding Remarks}\label{sec:concluding_remarks}

Energy-aware orchestration across geographically distributed infrastructures requires accounting for both the energy consumed by managed workloads and the \ac{GPU} energy cost of automated decision-making. This paper presented an agentic framework that measures these components jointly, introduced the \ac{NEB} to compare short-horizon infrastructure-energy changes with \ac{GPU} decision energy, and evaluated three reasoning configurations on a real multi-site Kubernetes testbed. For the evaluated lightweight workload and one-second attribution horizon, the observed infrastructure-energy differences were generally insufficient to offset the energy consumed by \ac{LLM} reasoning. ReAct and Plan~\&~Execute produced actionable tool calls, whereas the single-shot Tool~Calling configuration did not; prefix caching reduced mean ReAct \ac{GPU} energy by 29.4\%. These findings show that the sustainability of agentic orchestration depends jointly on actionability, reasoning strategy, workload magnitude, and the period over which operational savings are evaluated. Future work will investigate larger and more energy-intensive workloads, additional model families, adaptive strategies that match reasoning depth to intent complexity, longer post-action observation periods, and carbon-aware placement based on renewable-energy availability.


\section*{Acknowledgment}


The authors acknowledge support from FAPEMIG under Grant \#APQ00923-24; the FAPESP MCTIC/CGI Research Project SFI2 -- Slicing Future Internet Infrastructures under Grant 2018/23097-3; and Fundação para a Ciência e a Tecnologia (FCT) through the R\&D Unit Project UID/00319/2025, Centro ALGORITMI (ALGORITMI/UM). The authors also acknowledge support from the 6G-VERSUS project, funded by the Smart Networks and Services Joint Undertaking (SNS JU) under the European Union's Horizon Europe research and innovation programme under Grant Agreement No.~101192633.




\bibliographystyle{IEEEtran}
\bibliography{example}

@ARTICLE{paper1,
  author={Munir, Md. Shirajum and Tran, Nguyen H. and Saad, Walid and Hong, Choong Seon},
  journal={IEEE Transactions on Network and Service Management}, 
  title={{Multi-Agent Meta-Reinforcement Learning for Self-Powered and Sustainable Edge Computing Systems}}, 
  year={2021},
  volume={18},
  number={3},
  pages={3353-3374},
  doi={10.1109/TNSM.2021.3057960}}

@ARTICLE{paper2,
  author={Premsankar, Gopika and Ghaddar, Bissan},
  journal={IEEE Internet of Things Journal}, 
  title={{Energy-Efficient Service Placement for Latency-Sensitive Applications in Edge Computing}}, 
  year={2022},
  volume={9},
  number={18},
  pages={17926-17937},
  doi={10.1109/JIOT.2022.3162581}}

@article{paper3,
title = {{A forecasting-based perception layer for energy-aware resource management in LLM serving deployed on high-performance computing clusters}},
journal = {Sustainable Computing: Informatics and Systems},
volume = {50},
pages = {101333},
year = {2026},
issn = {2210-5379},
doi = {https://doi.org/10.1016/j.suscom.2026.101333},
url = {https://www.sciencedirect.com/science/article/pii/S2210537926000430},
author = {Jon Aguirre-Usandizaga and Ana I. Torre-Bastida and Miguel Linares De la puerta and Aitor Almeida}
}

@article{paper4, title={Energy-Aware Intelligent Computing Framework for Sustainable AI Workloads in Next-Generation Smart Systems}, volume={2}, url={https://submissions.adroidjournals.com/index.php/ijssic/article/view/201}, DOI={10.63503/j.ijssic.2025.201}, author={Kohli, Harshit and Hadi, Abdul and Mukhi, Nitin and Miah, Md Alamgir and Bushra Siddiqa, Kazi}, year={2026}, month={Jan.}, pages={34–47} }

@article{paper5,
author = {Kadhim, Mohammed Aboud and Hamad, Ahmed Rifaat and Ibraheem, Zeyid T.},
title = {{Advanced AI-Based Energy Efficiency and Network Performance Optimization in Next-Generation Wireless Systems Using Distributed Collective Intelligence}},
journal = {Journal of Electrical and Computer Engineering},
volume = {2026},
number = {1},
pages = {2275315},
doi = {https://doi.org/10.1155/jece/2275315},
url = {https://onlinelibrary.wiley.com/doi/abs/10.1155/jece/2275315},
eprint = {https://onlinelibrary.wiley.com/doi/pdf/10.1155/jece/2275315},
year = {2026}
}

@article{paper6, title={{AI-Driven Carbon-Aware Orchestration for Sustainable Cloud Deployments}}, volume={5}, url={http://ijeret.org/index.php/ijeret/article/view/555}, DOI={10.63282/3050-922X.IJERET-V5I2P117}, number={2}, journal={International Journal of Emerging Research in Engineering and Technology}, author={Suryadevara, Siva Sai Krishna}, year={2024}, month={Jun.}, pages={165–175} }

@ARTICLE{paper7,
  author={Zawish, Muhammad and Ashraf, Nouman and Ansari, Rafay Iqbal and Davy, Steven},
  journal={IEEE Internet of Things Journal}, 
  title={{Energy-Aware AI-Driven Framework for Edge-Computing-Based IoT Applications}}, 
  year={2023},
  volume={10},
  number={6},
  pages={5013-5023},
  doi={10.1109/JIOT.2022.3219202}}

@article{paper8, title={An Intelligent Optimization Model for Energy-Efficient Cloud Computing Systems}, volume={1}, url={https://csetx.org/index.php/cset/article/view/92}, DOI={10.66280/cset.v1i1.92}, number={1}, journal={Computer Science and Engineering Transactions}, author={Elena Vance}, year={2026}, month={Mar.} }

@article{paper9,
author = {Maisto, Vincenzo and Cilardo, Alessandro},
title = {{Compute-Efficient Modelling of Multi-NPU Inference on Edge MPSoCs for Energy-Aware Online Workload Allocation}},
year = {2026},
issue_date = {March 2026},
publisher = {Association for Computing Machinery},
address = {New York, NY, USA},
volume = {23},
number = {1},
issn = {1544-3566},
url = {https://doi.org/10.1145/3799238},
doi = {10.1145/3799238},
month = mar,
articleno = {33},
numpages = {25}
}

@inproceedings{paper10,
author = {Moore, Hayden and Qi, Sirui and Hogade, Ninad and Milojicic, Dejan and Bash, Cullen and Pasricha, Sudeep},
title = {{Sustainable Carbon-Aware and Water-Efficient LLM Scheduling in Geo-Distributed Cloud Datacenters}},
year = {2025},
isbn = {9798400714962},
publisher = {Association for Computing Machinery},
address = {New York, NY, USA},
url = {https://doi.org/10.1145/3716368.3735301},
doi = {10.1145/3716368.3735301},
booktitle = {Proceedings of the Great Lakes Symposium on VLSI 2025},
pages = {929–934},
numpages = {6},
location = {
},
series = {GLSVLSI '25}
}

@InProceedings{Andersen2026,
author="Andersen, Simon Malgo Pronk
and Mumberg, Laurits Christian Bang
and Golmohamadi, Hessam
and Albano, Michele",
editor="J{\o}rgensen, Bo N{\o}rregaard
and Ma, Zheng Grace
and Yussof, Salman",
title="Kubernetes Scheduling for Green-Powered Microgrid Data Centers",
booktitle="Energy Informatics",
year="2026",
publisher="Springer Nature Switzerland",
address="Cham",
pages="290--306",
isbn="978-3-032-19134-2"
}

@article{Danach2026,
author = {Danach, Kassem and Hamze, Kassem and Harb, Hassan and Kanj, Hassan},
title = {Carbon-Aware Scheduling in Cloud Computing Operations: A Multi-Objective Optimisation Approach},
journal = {IET Smart Grid},
volume = {9},
number = {1},
pages = {e70056},
doi = {https://doi.org/10.1049/stg2.70056},
url = {https://ietresearch.onlinelibrary.wiley.com/doi/abs/10.1049/stg2.70056},
eprint = {https://ietresearch.onlinelibrary.wiley.com/doi/pdf/10.1049/stg2.70056},
year = {2026}
}

@misc{Yang2025,
      title={A Survey on Task Scheduling in Carbon-Aware Container Orchestration}, 
      author={Jialin Yang and Zainab Saad and Jiajun Wu and Xiaoguang Niu and Henry Leung and Steve Drew},
      year={2025},
      eprint={2508.05949},
      archivePrefix={arXiv},
      primaryClass={cs.SE},
      url={https://arxiv.org/abs/2508.05949}, 
}

@ARTICLE{Bellal2026,
  author={Bellal, Zouhir and Lahlou, Laaziz and Kara, Nadjia and Murphy, Timothy and Phat Nguyen, Tan and Ahmed, Arif and Perez-Jimenez, Mario},
  journal={IEEE Transactions on Green Communications and Networking}, 
  title={Investigating the Potential of Kepler Toward Power Observability for Sustainable Cloud Computing}, 
  year={2026},
  volume={10},
  number={},
  pages={2118-2135},
  doi={10.1109/TGCN.2026.3660816}}

@article{Lee2026,
author = {Lee, Yubeen and Park, Eunil},
title = {{Toward Sustainable Agentic AI Systems: A Survey of Architectures and Methodologies}},
journal = {Sustainable Development},
volume = {n/a},
number = {n/a},
pages = {},
doi = {https://doi.org/10.1002/sd.70942},
url = {https://onlinelibrary.wiley.com/doi/abs/10.1002/sd.70942},
eprint = {https://onlinelibrary.wiley.com/doi/pdf/10.1002/sd.70942}
}

@INPROCEEDINGS{Mohamed2025,
  author={Mohamed, Laharraf and Hamza, Kamal Idrissi and Driss, Allaki},
  booktitle={2025 International Conference on Intelligent Systems: Theories and Applications (SITA)}, 
  title={{A Survey of Energy Efficiency for Kubernetes Clusters}}, 
  year={2025},
  volume={},
  number={},
  pages={1-6},
  doi={10.1109/SITA67914.2025.11273401}}

@Article{Jawaddi2024,
author={Agos Jawaddi, Siti Nuraishah
and Ismail, Azlan
and Sulaiman, Mohd Suffian
and Cardellini, Valeria},
title={{Analyzing Energy-Efficient and Kubernetes-Based Autoscaling of Microservices Using Probabilistic Model Checking}},
journal={Journal of Grid Computing},
year={2024},
month={Dec},
day={30},
volume={23},
number={1},
pages={3},
issn={1572-9184},
doi={10.1007/s10723-024-09789-9},
url={https://doi.org/10.1007/s10723-024-09789-9}
}

@Article{Kocot2023,
AUTHOR = {Kocot, Bartłomiej and Czarnul, Paweł and Proficz, Jerzy},
TITLE = {{Energy-Aware Scheduling for High-Performance Computing Systems: A Survey}},
JOURNAL = {Energies},
VOLUME = {16},
YEAR = {2023},
NUMBER = {2},
ARTICLE-NUMBER = {890},
URL = {https://www.mdpi.com/1996-1073/16/2/890},
ISSN = {1996-1073},
DOI = {10.3390/en16020890}
}

@article{Moreira2026,
title = {AI-driven orchestration at scale: Estimating service metrics on national-wide testbeds},
journal = {Future Generation Computer Systems},
volume = {174},
pages = {107971},
year = {2026},
issn = {0167-739X},
doi = {https://doi.org/10.1016/j.future.2025.107971},
url = {https://www.sciencedirect.com/science/article/pii/S0167739X25002663},
author = {Rodrigo Moreira and Rafael Pasquini and Joberto S.B. Martins and Tereza C. Carvalho and Flávio {de Oliveira Silva}}
}
%

\end{document}